\documentclass[twocolumn,floats,showpacs,superscriptaddress,pre]{revtex4}
\usepackage{amsmath,amssymb,bm}
\usepackage{graphics}
\usepackage{epsfig}
\usepackage{graphicx}

\begin{document}

\title{No-boundary Wave Functional and Own Mass of the Universe}
\author{Natalia Gorobey}
\affiliation{Peter the Great Saint Petersburg Polytechnic University, Polytekhnicheskaya
29, 195251, St. Petersburg, Russia}
\author{Alexander Lukyanenko}
\email{alex.lukyan@mail.ru}
\affiliation{Peter the Great Saint Petersburg Polytechnic University, Polytekhnicheskaya
29, 195251, St. Petersburg, Russia}
\author{A. V. Goltsev}
\affiliation{Ioffe Physical- Technical Institute, Polytekhnicheskaya 26, 195251, St.
Petersburg, Russia}

\begin{abstract}
An alternative formulation of the no-boundary initial state of the universe
in the Euclidean quantum theory of gravity is proposed. Unlike the
no-boundary Hartle-Hawking wave function, in which time appears together
with macroscopic space-time in the semiclassical approximation, in the
proposed formalism time is present from the very beginning on an equal
footing with spatial coordinates. The main element of the formalism is the
wave functional, which is defined on the world histories of the universe.
This ensures formal 4D covariance of the theory. The wave functional is
defined independently of the wave function as an eigenvector of the action
operator. The shape of the Origin region, together with the boundary
conditions, is determined by the structure of the total energy, which
includes the 3D invariant contribution of the expansion energy of the
universe with a minus sign. The proper mass of the universe arises as a
non-zero value of the expansion energy in the Origin and, over time, splits
into a spectrum of proper masses of 3D invariant dynamic modes. 4D
covariance is restored at zero own mass of the universe.
\end{abstract}

\maketitle







\section{\textbf{INTRODUCTION}}

The question of the origin of the universe has been and remains central to
cosmology. In this work we will focus on the idea of the quantum birth of
the universe from \textquotedblleft nothing\textquotedblright\ \cite{Vil,HH1,Lin,Rub,Vil2,ZelSta}.
It was most
consistently developed within the framework of the Euclidean quantum theory
of gravity (QTG) in the works of Hartle, Hawking, and Hertog \cite{HHH1}.
The main object in this approach is the representation of the no-boundary
wave function of the universe in the form of a functional integral
\begin{equation}
\psi =\int \prod Jdgd\varphi \exp \left( -\frac{1}{\hbar }\widetilde{I}%
_{GR}\right),  \label{1}
\end{equation}
where  $\widetilde{I}_{GR}$ is the action of General Relativity in Euclidean signature.
Integration is carried out over all Euclidean $4D$ metrics and
configurations of matter fields with given values on a single $3D$ boundary,
$J$ is the Faddeev-Popov determinant. However, in practice, when using polar
coordinates in the Origin \cite{HHH1}, integral Eq. (\ref{1}) is considered
as a representation of the Green's function for the Wheeler-De Witt (WDW)
equation with two boundary surfaces, one of which is contracted to a point -
a pole. In this case, it is not possible to completely get rid of the
boundary conditions for the fundamental dynamic variables at the pole. In
particular, the initial value of the scalar field remains a free parameter
\cite{HHH1}. A more aggravating circumstance is the fact that integral Eq.(%
\ref{1}) diverges and the no-boundary wave function can be given meaning
only within the framework of the semiclassical approximation. Therefore, in
the subsequent work \cite{HHH2}, the authors considered it reasonable to state
the problem in the semiclassical approximation without using a functional
integral, directly for the WDW equation, or through the holographic
principle \cite{HH2}. The reason for the divergence of integral Eq. (\ref{1})
is the uncertainty of the sign of the Hilbert-Einstein action. This problem
is closely related to the problem of the positivity of the gravitational
field energy \cite{H}. The latter was solved thanks to the proof of the
positive energy theorem for the case of asymptotically flat geometry
\cite{Witt,Fadd}. A modification of this theorem for the case of a closed
universe is considered in \cite{GL}. Here there is an irremovable negative
contribution to energy, which is entirely related to the expansion of the
universe.

This paper proposes a formalism alternative to the functional integral Eq. (%
\ref{1}) on the basis of the invariant wave functional $\Psi \left[ g\left(
x,t\right) ,\varphi \left( x,t\right) \right] $, which is defined on the
space of $4D$ world histories of the universe. To avoid terminological
confusion, we immediately emphasize that the wave function $\psi \left(
g_{ik}\left( x\right) ,\varphi \left( x\right) ,N,N^{k},t\right) $ is a
functional of the functions $g_{ik}\left( x\right) ,\varphi \left( x\right) $
on a $3D$ spatial section at a given time $t$ and a functional of the given
lapse and shift functions $N,N^{k}$ \cite{MTW}. To determine the wave
functional, the work \cite{GLG} formulated the quantum principle of least
action, according to which the wave functional is an eigenvector of the
action operator.

In the new formalism, the integration over $N,N^{k}$ is initially absent. In
the covariant quantum theory, based on the Batalin-Fradkin-Vilkovysky theorem
\cite{BatVil,FradVil}, the integration over the lapse function $N$ is
equivalent to the integration over proper time (see \cite{Gov}), so in the new
formalism time remains a free parameter. This makes it possible to formulate
a boundary value problem for the wave functional in the \textquotedblleft
subpolar\textquotedblright\ region (Euclidean instanton), in which the pole
is an internal point, without any additional conditions for the fundamental
dynamic variables in it. To fix time in an instanton, one additional
parameter will be required - the own mass of the universe.

The next section formulates the basic concepts of the canonical formalism
and a new description of the dynamics in quantum theory of gravity. The
second section gives a representation of the energy of a closed universe
using spin variables. In the third section, the boundary value problem for
the Euclidean instanton is considered in the case of a homogeneous isotropic
model of the universe, in which the concept of its own mass arises. In the
fourth section, a new canonical representation of the action of the theory
of gravity is introduced, based on the energy structure of a closed
universe, in which the own mass is realized in the form of a mass spectrum
of individual $3D$ invariant dynamic modes.

\section{WAVE FUNCTIONAL IN THE QUANTUM THEORY OF GRAVITY}

Let's start our consideration with the classical action of general relativity%
\begin{equation}
I_{GR}=-\frac{1}{16\pi G}\int \sqrt{-g}d^{4}xR+I_{m}\left[ g,\varphi \right]
.  \label{2}
\end{equation}%
Using $3+1$ splitting of the metric

\begin{equation}
ds^{2}=\left( Ndt\right) ^{2}-g_{ik}\left( dx^{i}+N^{i}dt\right) \left(
dx^{k}+N^{k}dt\right) ,  \label{3}
\end{equation}%
let's write it in the canonical form of Arnovitt, Deser, and Misner (ADM)
\cite{ADM}:

\begin{equation}
I_{ADM}=\int dt\int_{\Sigma }d^{3}x\left( \overset{\cdot }{g}_{ik}\pi ^{ik}-N%
\mathcal{H-}N_{i}\mathcal{H}^{i}\right) ,  \label{4}
\end{equation}%
$N_{i}=g_{ik}N^{k}$ , where
\begin{equation}
\mathcal{H}\left( \pi ^{ik},g_{ik}\right) =-\frac{1}{\sqrt{g}}\left[ Tr%
\mathbf{\pi }^{2}-\left( Tr\mathbf{\pi }\right) ^{2}\right] +\sqrt{g}R,+%
\mathcal{H}_{m},  \label{5}
\end{equation}%
\begin{equation}
\mathcal{H}^{i}=2\pi _{\left\vert k\right. }^{ik}+\mathcal{H}_{m}^{i}
\label{6}
\end{equation}%
are Hamiltonian and momentum constraints. and the canonical momenta
conjugated to the $3D$ metric tensor $g_{ik}$ have the form:
\begin{equation}
\pi ^{ik}=\sqrt{g^{\left( 3\right) }}\left( g^{ik}Tr\mathbf{K-}K^{ik}\right)
,  \label{7}
\end{equation}
\begin{equation}
K_{ik}=\frac{1}{2N}\left( N_{i\left\vert k\right. }+N_{k\left\vert i\right.
}-\frac{\partial g_{ik}}{\partial t}\right) .  \label{8}
\end{equation}%
The last terms in Eqs. (\ref{5}) and (\ref{6}) are the energy
and momentum of the matter fields, respectively.

In order to describe the evolution of the universe in QTG in terms of world
histories, we introduce the state functional $\Psi $. We define it as the
product of wave functions $\psi \left( g_{ik}\left( x\right) ,\varphi \left(
x\right) ,N,N^{k},t\right) $ on spatial sections $\Sigma _{n}$ for each time
$t_{n}=\varepsilon n,\varepsilon =T/n$. We suppose that the time dependence
of the wave function is determined by the Schr\"{o}dinger equation
\begin{equation}
i\hbar \frac{\partial \psi }{\partial t}=\int_{\Sigma }d^{3}x\left( N%
\widehat{\mathcal{H}}+N_{k}\widehat{\mathcal{H}}^{k}\right) \psi ,  \label{9}
\end{equation}%
where the lapse and shift functions $N,N_{k}$ are considered fixed.
Consequently, the wave function $\psi $ is also a functional of $N,N_{k}$,
and the WDW wave equations
\begin{equation}
\widehat{\mathcal{H}}\psi =\widehat{\mathcal{H}}^{i}\psi =0  \label{10}
\end{equation}%
are not initially postulated, which means they may not be fulfilled. For the
wave functional $\Psi $, the normalization condition is assumed to be
satisfied:
\begin{equation}
\left\langle \Psi \right\vert \left. \Psi \right\rangle =\int \prod
Jdgd\varphi \overline{\Psi }\Psi .  \label{11}
\end{equation}
It should be assumed that, being a functional of $4D$ geometry (including
the lapse and shift functions $N,N_{k}$), the wave functional is an
invariant of general covariant transformations. The assumption is based on
the fact that the basic equation of motion - the Schr\"{o}dinger equation
Eq. (\ref{9}) for the wave function $\psi $ can be equivalently replaced by
the corresponding equation for the wave functional $\Psi $. The latter is a
secular equation for the action operator, which is obtained by quantizing
direct the action of ADM Eq. (\ref{4}) \cite{GLG}. This means that we have
the opportunity to calculate, for example, the average values of expressions
containing the first and second derivatives with respect to time, in
particular,
\begin{equation}
\left\langle \Psi \right\vert R_{\mu \nu }\left\vert \Psi \right\rangle ,
\label{12}
\end{equation}%
where $R_{\mu \nu }$ is the $4D$ Ricci tensor. Based on the above, we should
expect that expression Eq. (\ref{12}) forms a tensor of the second rank with
respect to arbitrary transformations of space-time coordinates, as in the
classical theory. In particular, it is easy to check that the average values
of both sides of relations Eqs. (\ref{7}) and (\ref{8}), where

\begin{equation}
\pi ^{ik}\rightarrow \widehat{\pi }^{ik}\left( x\right) =\frac{\hslash }{i}%
\frac{\delta }{\delta g_{ik}\left( x\right) },  \label{13}
\end{equation}%
obey the same equalities after quantization. As we will see, this cannot be
said for the constraint equations.

\section{THE ENERGY OF A CLOSED UNIVERSE}

The lapse and shift functions $N,N_{k}$ in the new formalism remain
arbitrary, integration over them is carried out only under the normalization
condition Eq. (\ref{11}). Next, we will introduce a special spin
parametrization of these functions, and at the same time the Ashtekar \cite%
{Ash1} complex representation of canonical variables of the gravitational
field ($\widetilde{\sigma }_{AB}^{k},A_{KAB}$, $A,B=0,1$ - spin indices). We
immediately take into account the so-called reality condition for the
Ashtekar connection, setting
\begin{equation}
A_{kAB}=\Gamma _{kAB}\left( \sigma \right) +\frac{i}{\sqrt{2}}M_{kAB},
\label{14}
\end{equation}%
where $\Gamma _{kAB}\left( \sigma \right) $ are components of the real
spin-connection, and $M_{kAB}$ are the canonical momenta conjugated to the
spin variables $\widetilde{\sigma }_{AB}^{k}$ in the real representation, in
which we can also immediately put
\begin{equation}
M_{kAB}=\frac{\pi _{kl}\sigma _{AB}^{l}}{\sqrt{g^{\left( 3\right) }}}
\label{15}
\end{equation}%
(Gaussian constraint $\mathcal{P}^{AB}$ of Ashtekar). Let us introduce the $%
3D$ Dirac operator on a spatial section $\Sigma $:
\begin{equation}
\mathcal{D}\eta \equiv i\sqrt{2}\binom{n_{A^{\prime }{}^{\prime
}}^{A^{\prime }}\overline{\sigma }_{B^{\prime }}^{kA^{\prime }}\overline{%
\triangledown }_{k}\overline{\mu }^{B^{\prime }}}{n_{A}^{A^{\prime }}\sigma
_{B}^{kA}\triangledown _{k}\lambda ^{B}},  \label{16}
\end{equation}%
where $\eta $ is the bispinor Dirac field on the spatial section $\Sigma $,
\begin{equation}
\eta =\binom{\lambda ^{A}}{\overline{\mu }^{A^{\prime }}},  \label{17}
\end{equation}%
and $n_{A^{\prime }{}^{\prime }}^{A^{\prime }}$is an arbitrary unitary
matrix (spin-tensor) in the spin space. The complex covariant derivative of
a spinor field is defined as follows:
\begin{equation}
\triangledown _{k}\lambda _{A}\equiv \partial _{k}\lambda
_{A}+A_{kA}^{B}\lambda _{B}.  \label{18}
\end{equation}%
Let us introduce anti-involution in the spin space,
\begin{equation}
\lambda _{A}^{+}\equiv \sqrt{2}n_{A^{\prime }{}^{\prime }}^{A^{\prime }}%
\overline{\lambda }^{A^{\prime }},\left( \lambda _{A}^{++}=-\lambda
_{A}\right) .  \label{19}
\end{equation}%
We assume that $\sigma _{AB}^{k+}=\sigma _{AB}^{k}$. Let us also introduce
the Hermitian scalar product in the spin space:
\begin{equation}
\left( \eta _{1},\eta _{2}\right) \equiv \int_{\Sigma }\sqrt{g^{\left(
3\right) }}d^{3}xn_{AA^{\prime }}\left( \lambda _{1}^{A}\overline{\lambda }%
_{2}^{A^{\prime }}+\overline{\mu }_{1}^{A^{\prime }}\mu _{2}^{A}\right) .
\label{20}
\end{equation}%
It is easy to verify that the Dirac operator Eq. (\ref{16}) is Hermitian
with respect to this scalar product.
Our constructions are based on the Witten identity, which relates the
difference of two positive definite quadratic forms of the bispinor $\eta $
with a linear combination of gravitational constraints in the Ashtekar
representation (see, \cite{GLG}),
\begin{eqnarray}
\left( \eta ,W\eta \right) &\equiv &-\frac{11}{9}\left( \eta ,\mathcal{D}%
^{2}\eta \right) +\left( \eta ,\left( -\Delta +\mathcal{T}\right) \eta
\right)  \notag \\
&\equiv &\mathcal{L}\left( \widetilde{C},\widetilde{C}_{k},\widetilde{%
\mathcal{P}}^{AB}\right) ,  \label{21}
\end{eqnarray}%
The coefficients of the linear combination are the lapse and shift
functions, as well as the zero components of the Ashtekar connection of the
form \cite{Ash2}:
\begin{equation}
N=\frac{1}{8}n_{AA^{\prime }}\left( \lambda ^{A}\overline{\lambda }%
^{A^{\prime }}+\overline{\mu }^{A^{\prime }}\mu ^{A}\right) ,  \label{22}
\end{equation}
\begin{equation}
N^{k}=-\frac{i}{4}\sigma _{AB}^{k}\left( \lambda ^{A}\lambda ^{+B}+\mu
^{+A}\mu ^{B}\right) ,  \label{23}
\end{equation}
\begin{equation}
A_{0AB}=-\frac{1}{16\sqrt{2}}\sigma _{\left( A\right\vert \left.
C\right\vert }^{m}\left( \triangledown _{m}\lambda _{\left. B\right)
}\lambda ^{+C}+\triangledown _{m}\mu _{\left. B\right) }\mu ^{+C}\right) .
\label{24}
\end{equation}%
The second term in Eq.(\ref{21}) has the form:
\begin{eqnarray}
&&\left( \eta ,\left( -\Delta +\mathcal{T}\right) \eta \right)  \notag \\
&=&\frac{1}{2}\int_{\Sigma }\sqrt{g^{\left( 3\right) }}d^{3}xn_{AA^{\prime
}}n_{MM^{\prime }}n_{NN^{\prime }}\left( \xi ^{AMN}\overline{\xi }%
^{A^{\prime }M^{\prime }N^{\prime }}\right.  \notag \\
&&\left. +\chi ^{AMN}\overline{\chi }^{A^{\prime }M^{\prime }N^{\prime
}}\right) +\left( \eta ,\mathcal{T}\eta \right) ,  \label{25}
\end{eqnarray}%
where
\begin{equation}
\chi ^{MNA}\equiv \sigma ^{mMN}\triangledown _{m}\mu ^{A}+\frac{2}{3}%
\epsilon ^{A\left( M\right. }\sigma _{P}^{m\left. N\right) }\triangledown
_{m}\mu ^{P},  \label{26}
\end{equation}
\begin{equation}
\xi ^{MNA}\equiv \sigma ^{mMN}\triangledown _{m}\lambda ^{A}+\frac{2}{3}%
\epsilon ^{A\left( M\right. }\sigma _{P}^{m\left. N\right) }\triangledown
_{m}\lambda \mu ^{P},  \label{27}
\end{equation}%
where $\epsilon ^{AB}$ is a completely antisymmetric unit spin tensor. Spin
tensors Eqs. (\ref{26})  and (\ref{27}) are completely symmetric. The last term
on the right side of Eq. (\ref{25}) is a positive definite form of the
energy-momentum tensor of matter fields. Thus, identity Eq. (\ref{21}) gives
a representation of the Hamilton function of the theory of gravity
(right-hand side of Eq. (\ref{21})) as the difference of two positive
definite quadratic forms of the bispinor $\eta $. The fact that we thus
obtain the Hamilton function in an arbitrary gauge follows from counting the
number of real constraints of the theory of gravity (7 pieces) and the
number of independent real parameters of the bispinor $\eta $ (8 pieces).
The presence of a redundant parameter means the degeneracy of the quadratic
form of the operator
\begin{equation}
W=-\frac{11}{9}\mathcal{D}^{2}+\left( -\Delta +\mathcal{T}\right) ,
\label{28}
\end{equation}%
i.e., the existence of a zero eigenvalue for this operator.

In the representation of the Hamilton function of a closed universe Eq.(\ref%
{21}), separation of the contributions of energy components with different
signs has been achieved. The quadratic form $\left( \eta ,\mathcal{D}%
^{2}\eta \right) $ contains the kinetic energy $\left( Tr\mathbf{\pi }%
\right) ^{2}$ (together with the corresponding potential energy), it
describes the dynamics of the $3D$ geometry scale factor $\sqrt{g^{\left(
3\right) }}$. Therefore, we will call it the energy of space. The quadratic
form $\left( \eta ,\Delta \eta \right) $ does not contain $\left( Tr\mathbf{%
\pi }\right) ^{2}$, and describes the dynamics of the \textquotedblleft
transverse\textquotedblright\ components of the gravitational field that
describe gravitational waves. We will call it, together with $\left( \eta ,%
\mathcal{T}\eta \right) $, the energy of matter. The explicit separation of
these two components in Eq. (\ref{21}) is a version of the positive energy
(of matter) theorem for the case of a closed universe. The combination of
signs in Eq. (\ref{21}) also determines the signature of the configuration
space of the theory of gravity (superspace).

We can now discuss the issue of regularizing the convergence of the
functional integral representation of the kernel of the evolution operator
for the Schr\"{o}dinger equation Eq. (\ref{9}). For the functional integral
to converge, it is necessary that the total energy of the universe have a
certain sign. This can be achieved by introducing a variable value $e$
instead of the minus sign in Eq. (\ref{21}), which is taken equal to $+1$ at
the calculation stage. In the already found wave function, along with the
return to real time, the sign of $e$ should also be changed.

\section{EUCLIDEAN BEGINNING OF A HOMOGENEOUS ISOTROPIC MODEL OF THE UNIVERSE%
}

The transition to describing the quantum evolution of the universe in terms
of world histories and wave functional allows us to take a fresh look at the
problem of initial data for this evolution. In the classical theory of
gravity, the time lines of the universe begin at one point, which is the Big
Bang singularity. In Euclidean QTG, these lines simply serve as meridians of
the \textquotedblleft polar\textquotedblright\ coordinate system \cite{HHH1}%
. The pole itself has no features other than a coordinate singularity.
Therefore, in \cite{GLG}, the state of the universe in the \textquotedblleft
subpolar\textquotedblright\ region (with one boundary -- along the
\textquotedblleft polar\textquotedblright\ circle) was proposed to be sought
in a non-singular coordinate system using the generalized canonical
De-Donder-Weil formalism. And although to introduce time we return to the
usual $3+1$ ADM splitting of the metric in polar coordinates, at the pole
itself, as an equal point, we place not the initial data for the fundamental
dynamic variables $\left( g,\varphi \right) $, but their distribution in
terms of the wave functional $\Psi \left[ g,\varphi \right] $. In this
sense, we call the wave functional of the universe no-boundary.

Let us consider in more detail the initial stage of evolution of the
homogeneous isotropic Friedmann-Lemaitre universe with the metric
\begin{equation}
ds^{2}=N^{2}\left( t\right) dt^{2}-a^{2}\left( t\right) d\Omega _{3}^{2},
\label{29}
\end{equation}%
where $d\Omega _{3}^{2}$ is an element of length on a $3D$ sphere of unit
radius, with a real scalar field and zero cosmological constant. Its
dynamics are described by the action (Lorentzian signature)
\begin{eqnarray}
I_{FL}\left[ a,\phi \right] &=&\frac{1}{2}\int_{0}^{T}dt\left[ -\frac{a}{%
\gamma }\left( \frac{\overset{\cdot }{a}^{2}}{N}-N\right) \right.  \notag \\
&&\left. +2\pi ^{2}a^{3}\left( \frac{\overset{\cdot }{\phi }^{2}}{N}-V\left(
\phi \right) N\right) \right] ,  \label{30}
\end{eqnarray}%
where $\gamma =2G/3\pi $. The Hamilton function and the corresponding Schr%
\"{o}dinger equation for this model are:
\begin{eqnarray}
h_{FL} &=&N\mathcal{H}_{FL}=N\frac{1}{2}\left[ -\left( \frac{\gamma p_{a}^{2}%
}{a}+a\right) \right.  \notag \\
&&\left. +\left( \frac{p_{\phi }^{2}}{2\pi ^{2}a^{3}}+2\pi ^{2}a^{3}V\left(
\phi \right) \right) \right] ,  \label{31}
\end{eqnarray}
\begin{equation}
i\hbar \frac{\partial \psi }{\partial s}=\widehat{\mathcal{H}}_{FL}\psi
,s=\int_{0}^{t}N\left( t\right) dt.  \label{32}
\end{equation}
We will further restrict ourselves to the semiclassical approximation;
therefore, we do not consider the problem of ordering noncommuting factors
in $\widehat{\mathcal{H}}_{FL}$ here. We also do not consider the problem of
convergence of the Euclidean functional integral, which represents the
kernel of the evolution operator for equation Eq. (\ref{32}). The extremum
conditions for the Euclidean action, which is obtained from Eq. (\ref{30})
after the transition to imaginary time $s=-i\tau $, have the form:
\begin{equation}
a\overset{\cdot \cdot }{a}+\frac{1}{2}\left( \overset{\cdot }{a}%
^{2}-1\right) -3\pi ^{2}a^{2}\left( \overset{\cdot }{\phi }^{2}+V\right) =0
\label{33}
\end{equation}%
is the extremum condition in $a$ and
\begin{equation}
\overset{\cdot \cdot }{\phi }+3\frac{\overset{\cdot }{a}}{a}\overset{\cdot }{%
\phi }-V^{\prime }\left( \phi \right) =0  \label{34}
\end{equation}%
is extremum condition with respect to $\phi $, where the dot denotes the
derivative with respect to $\tau ,\tau \in \left[ 0,T\right] $. Let us
immediately note that the constraint equation $\mathcal{H}_{FL}=0$ is not
among the extremum conditions, since the lapse function $N$ is not
considered as a dynamic variable, and the integral of it is the proper time $%
s$.

Now let's consider the problem of boundary conditions for differential
equations Eqs. (\ref{33}), (\ref{34}). In \cite{HHH1}, the Euclidean
functional integral of the form Eq. (\ref{1}) is taken on a compact region
of $4D$ Riemannian space with a single boundary on which the values of the
scale factor $a(T)=b$ and the scalar field $\phi (T)=\chi $ are given. At
the \textquotedblleft pole\textquotedblright\ \textquotedblleft
natural\textquotedblright\ initial conditions are chosen
\begin{equation}
a\left( 0\right) =0,\overset{\cdot }{\phi }\left( 0\right) =0.  \label{35}
\end{equation}%
However, the composition of the equations -- extremum conditions in the work
\cite{HHH1} differs from Eqs. (\ref{33}) and (\ref{34}). Since integral Eq. (\ref%
{1}) contains additional integration over proper time, the constraint
equation also arises under extremum conditions. And since the constraint is
also the first integral of the equations of motion Eqs. (\ref{33}) and (\ref{34}%
), one of them, namely equation Eq. (\ref{34}), can be considered redundant.
With this formulation of the boundary value problem, the free parameter
turns out to be the value of the scalar field at the pole $\phi \left(
0\right) =\phi _{0}$. But this contradicts the very idea of constructing a
no-boundary wave function, which assumes the absence of any initial data for
fundamental dynamic variables in the polar region. This does not apply to
conditions Eq. (\ref{35}), which arise precisely as a result of the choice of
a polar coordinate system in a homogeneous isotropic model of the universe.

Let's see how the second of the \textquotedblleft natural\textquotedblright\
conditions Eq. (\ref{35}) arises if we consider it as the primary
representation of the evolution operator in non-singular coordinates in the
subpolar region. Moving along the meridian to the pole (one of the time
lines in polar coordinates), beyond the pole we will smoothly continue this
movement along the opposite (at an angle $180^{0}$) meridian, connecting
them into one time line of a non-singular coordinate grid. Let us divide
this time axis into small sections of length $\varepsilon $ and write the
contribution of the scalar field to the functional integral for the
evolution operator of the pole and neighboring points located symmetrically:
\begin{eqnarray}
&&\int ...d\phi _{0}...\exp \left\{ -\frac{1}{\hbar }\pi ^{2}\left[ \left(
\frac{a_{-1}}{2}\right) ^{3}\right. \right.  \notag \\
&&\times \left( \frac{\left( \phi _{0}-\phi _{-1}\right) ^{2}}{\varepsilon }%
+V\left( \frac{\phi _{0}+\phi _{-1}}{2}\right) \varepsilon \right)  \notag \\
&&\left. \left. +\left( \frac{a_{1}}{2}\right) ^{3}\left( \frac{\left( \phi
_{0}-\phi _{1}\right) ^{2}}{\varepsilon }+V\left( \frac{\phi _{0}+\phi _{1}}{%
2}\right) \varepsilon \right) \right] \right\}  \label{36}
\end{eqnarray}%
To calculate this integral using the steepest descent method, we find the
extremum of the exponent in $\phi _{0}$, which (in the limit $\varepsilon
\rightarrow 0$) gives: $\phi _{0}=\phi _{1}$. Here we also took into account
the symmetry of the model under consideration, $\phi _{1}=\phi _{-1}$, $%
a_{1}=a_{-1}$. Thus, the second condition in Eq. (\ref{35}) arises as a
consequence of estimating the integral over $\phi _{0}$ in the functional
integral representation of the propagator. The presence of this integral
also means that the initial condition for the wave function at the pole (at $%
\tau =0$) should be taken
\begin{equation}
\psi _{0}=A\delta \left( a\right) .  \label{37}
\end{equation}%
Thus, natural initial conditions mean that initially $a=0$, and the field $%
\phi $ can take on any value with equal probability.

To complete the formulation of the boundary problem, we define the boundary
conditions at $\tau =T$. Equations (\ref{33}) and (\ref{34}) determine the
initial instanton in the Euclidean region if its right boundary point on the
$a$-axis is a cusp point, i.e.,
\begin{equation}
\overset{\cdot }{a}\left( T\right) =0.  \label{38}
\end{equation}%
Thus, the history of the scale factor $a\left( \tau \right) $ in the
instanton is completely determined. For a given $T$, the history of the
scalar field $\phi \left( \tau \right) $, including its initial $\phi _{0}$
(as well as final $\phi \left( T\right) $) value, also becomes completely
determined, since the shape of the potential well for the instanton $a\left(
\tau \right) $ is determined by the function $\phi \left( \tau \right) $.
There remains one undefined parameter $T$, fixed by us. We can still
calculate the first integral of the equations of motion, which in the
general case is constant, but not equal to zero:
\begin{equation}
\mathcal{H}_{FL}\left( \tau \right) =-M^{2}\neq 0.  \label{39}
\end{equation}%
As we remember, the constraint equation $\mathcal{H}_{FL}=0$ serves
precisely to determine the time of movement $T$ in the generally accepted
approach. However, here this constraint equation, in the presence of a free
time parameter, does not follow from anywhere, and we are forced to accept
as an additional possibility the presence of a non-zero own mass of the
universe $M^{2}$ in Eq. (\ref{39}). The result can be formulated
differently: if the own mass of the universe is given, the shape of the
initial instanton in the Euclidean QTG with its own time is completely
determined. The minus sign in Eq. (\ref{39}) follows from the analysis of the
asymptotic behavior of the scale factor at the pole. It is easy to check that
\begin{eqnarray}
&&a\symbol{126}\left( \frac{9}{2}\right) ^{1/3}M^{2/3}\tau ^{2/3}  \notag \\
&&+\frac{9}{20M^{2/3}}\left( \frac{2}{9}\right) ^{1/3}\tau ^{4/3}+...
\label{40}
\end{eqnarray}%
at $\tau \rightarrow 0$. Thus, the spatial part of the energy of the
universe dominates in the Beginning, and this serves as a source of its
expansion. Simple asymptotic behavior Eq. (\ref{40}) and the entire
expansion picture will change if we also take into account the dynamics of
anisotropy near the beginning of \cite{MTW}. However, the main term in
asymptotics Eq. (\ref{40}) will be preserved, as well as the meaning of the
constant $M$. The proper mass remains constant only in a homogeneous
isotropic model of the universe. In general, this is not the case, and the
dynamics of one's own mass can be directly related to the universe's own
time.

\section{OWN MASS AND PROPER TIME IN AN INHOMOGENEOUS UNIVERSE}

To establish the connection between proper mass and proper time in the
general case, let us consider the new canonical representation of the theory
of gravity, which is naturally induced by the representation of the Hamilton
function Eq. (\ref{21}). If we consider the bispinor $\eta $ as an
independent dynamic variable, then the corresponding Euler-Lagrange equation
has the form:
\begin{equation}
W\eta =0.  \label{41}
\end{equation}%
Taking into account that $\eta $ is initially considered as an arbitrary
bi-spinor, we obtain a representation of the system of gravitational
connections in the form of an operator equation
\begin{equation}
W=0.  \label{42}
\end{equation}%
The operator $W$ is Hermitian on the space of bispinors and its spectrum is
real. The operator itself is equal to zero if and only if all its
eigenvalues $w_{n}$ are equal to zero. The eigenvalues, as well as the
eigenvectors $\eta _{n}$, are functions of the fundamental canonical
variables. The eigenvalues $w_{n}$ form a closed algebra with respect to
Poisson brackets:
\begin{equation}
\left\{ w_{n},w_{m}\right\} =C_{nm}^{p}w_{p},  \label{43}
\end{equation}%
in which the structural \textquotedblleft constants\textquotedblright\ $%
C_{nm}^{p}$ are determined by the eigenvectors $\eta _{n}$, i.e. are also
functions of canonical variables. We will further call eigenvalues $w_{n}$
as dynamic modes. Expanding an arbitrary bispinor $\eta $ over a complete
(orthonormal) set of eigenfunctions,
\begin{equation}
\eta =\sum_{n}\zeta ^{n}\eta _{n},  \label{44}
\end{equation}%
we can represent the Hamilton function of gravity theory as a linear
combination of a new set of constraints:
\begin{equation}
\left( \eta ,W\eta \right) =\sum_{n}L^{n}w_{n},L^{n}=\left\vert \zeta
^{n}\right\vert ^{2}.  \label{45}
\end{equation}%
Arbitrary Lagrange multipliers $L^{n}$ under infinitesimal general covariant
transformations generated by $w_{n}$ constraints,
\begin{equation}
\delta A=\delta s^{m}\left\{ A,w_{m}\right\} ,  \label{46}
\end{equation}%
where $A$ is an arbitrary function of canonical variables, must be
transformed as follows
\begin{equation}
\delta L^{n}=\delta \overset{\cdot }{s}^{n}-C_{mp}^{n}L^{m}\delta s^{p}
\label{47}
\end{equation}%
to ensure action invariance. These infinitesimal transformations are
generated by infinitesimal shifts of the proper time parameters $s^{n}$, and
the generators of these shifts are the eigenvalues $w_{n}$. To determine the
Lagrange multipliers corresponding to finite values of the proper time
parameters, equation Eq. (\ref{47}) can be solved iteratively, and the
solution can be represented as a power series:
\begin{equation}
L^{m}=\Lambda _{n}^{m}\left( s\right) \overset{\cdot }{s}^{n},  \label{48}
\end{equation}
\begin{eqnarray}
\Lambda _{n}^{m}\left( s\right) &=&\delta _{n}^{m}-C_{np}^{m}s^{p}  \notag \\
&&+\frac{1}{2!}C_{rp}^{m}C_{nq}^{r}s^{p}s^{q}+....  \label{49}
\end{eqnarray}%
The proper time parameters introduced in this way are integrals of the
Lagrange multipliers:
\begin{equation}
\int_{0}^{T}dtL^{m}\left( t\right) =\int_{0}^{S^{n}}\Lambda _{n}^{m}\left(
s,C\right) ds^{n}.  \label{50}
\end{equation}%
The values of the canonical variables in the structure functions $C_{np}^{m}$
are taken at the same moment of coordinate time $t$ as the proper time
parameters $s^{p}$. The time evolution of the eigenvalues $w_{n}$ is
determined by the equations
\begin{eqnarray}
\frac{dw_{n}}{dt} &=&\frac{\partial w_{n}}{\partial s^{p}}\overset{\cdot }{s}%
^{p}=\left\{ w_{n},L^{m}w_{m}\right\}  \notag \\
&=&\left\{ w_{n},\Lambda _{p}^{m}\right\} \overset{\cdot }{s}^{p}+\Lambda
_{p}^{m}C_{nm}^{q}w_{q}\overset{\cdot }{s}^{p},  \label{51}
\end{eqnarray}%
i.e.
\begin{equation}
\frac{\partial w_{n}}{\partial s^{p}}=\left\{ w_{n},\Lambda _{p}^{m}\right\}
+\Lambda _{p}^{m}C_{nm}^{q}w_{q}.  \label{52}
\end{equation}%
In quantum theory, all these relations should be considered in the form of
average values in the state described by the wave functional $\Psi $. It
follows that if the eigenvalues w$_{n}$ are zero at the beginning (classical
constraints), they always remain so. In this case, we can talk about
preserving the $4D$ covariance of the theory. If at first there is a
non-zero intrinsic mass in some dynamic mode,
\begin{equation}
w_{n}=-m_{n}^{2}\neq 0,  \label{53}
\end{equation}%
the distribution of own masses over modes will change over time, and this
change itself can be considered as a measure of proper time.

Thus, the Euclidean instanton in the general case has the following
structure in polar coordinates (radial coordinate - time axis). At the pole
(approaching the pole), the approximation of a homogeneous, isotropic model
of the universe with a single dynamic mode described by the Hamilton
function $\mathcal{H}_{FL}$ is valid. This will happen when choosing polar
coordinates in a small neighborhood of any interior point of a smooth
manifold. Accordingly, this dynamic mode can be associated with its own mass
$M$ as the only parameter of the universe model. The Euclidean
\textquotedblleft evolution\textquotedblright\ of the instanton along the
radial axes is given by the equation
\begin{equation}
\frac{d}{dt}\sqrt{g^{\left( 3\right) }}=\left\{ \sqrt{g^{\left( 3\right) }}%
,L^{m}w_{m}\right\} .  \label{54}
\end{equation}%
We actually have an infinite set of equations (one for each point of the
spatial section). The spatial boundary of the Euclidean instanton is
determined by the condition that the derivative of $\sqrt{g^{\left( 3\right)
}}$ with respect to time is equal to zero at all spatial points. This gives
a system of equations for determining the complete set of proper time
parameters at the boundary, and the system of equations Eq. (\ref{52})
allows us to find the resulting distribution of proper mass over modes.

\section{CONCLUSIONS}

The generally accepted formulation of the covariant quantum theory of
gravity, based on the WDW equations, as well as using the formalism of the
invariant functional integral, gives rise to the problem of time (more
precisely, its absence). Along with time, the possibility of introducing any
additional quantities, in addition to the set of fundamental dynamic
variables and associated parameters of the original Lagrangian, is excluded.
However, the observed evolution of the universe (or the generally accepted
interpretation of observational data) and the idea of the Big Bang as the
beginning of this evolution, one way or another, require the introduction of
time. This can be achieved by identifying the time parameter with a suitable
fundamental dynamic variable \cite{Ash3}. In this case, time acquires a
material character in the literal sense of the word, if one of the fields of
matter is taken as such a variable. In this paper, an alternative option is
proposed - the preservation of the coordinate time parameter of the
classical theory of gravity in quantum theory. This is achieved by
transitioning the description of the quantum state of the universe from a $%
3D $ distribution on a spatial section $\Sigma $ to a description in terms
of the wave functional on $4D$ world histories. With this modification, the
formal covariance of quantum theory is preserved in the same form as in the
classical one, when time and spatial coordinates were equal. However, this
equality is actually violated in the case of a closed universe by the
signature of the configuration space: the negative contribution in it is
clearly highlighted by the $3D$ invariant quadratic form of the expansion
energy, corresponding to the degrees of freedom of the scale factor $\sqrt{%
g^{\left( 3\right) }}$. This energy structure of the universe determines the
shape of the initial Euclidean instanton in the semiclassical approximation.
We also associate with this $3D$ invariant energy structure the spectrum of
parameters of the proper time and the canonically conjugate spectrum of
parameters of the own mass of the universe. If the proper mass, the
distribution and motion of which in space can be associated with a selected
reference frame, is assumed to be equal to zero, there is no physical reason
for the violation of the $4D$ covariance of the theory. General covariance
can be preserved even with a non-zero own mass if it is a constant of
motion. But this is possible when the structure constants in Eq. (\ref{43})
are equal to zero, i.e., the dynamic modes in the theory of gravity are
completely independent.

\section{ACKNOWLEDGEMENTS}

We are thanks V.A. Franke for useful discussions.





\bigskip

\begin{thebibliography}{99}
\bibitem{Vil} Vilenkin, A. Creating of the Universese from Nothing. \textit{%
Phys. Lett}. 1982, 117B, 25-28.

\bibitem{HH1} Hartle, J. B.; Hawking, S.W. Wave function of the Universe.
\textit{Phys. Rev}. D 1983, 28, 2960.

\bibitem{Lin} Linde, A.D. (1984). Quantum creation of the inflationary
universe, \textit{Lett. Nuovo Cimento} 39, 401.

\bibitem{Rub} Rubakov, V.A. (1984). Quantum mechanics in the tunneling
universe, \textit{Phys. Lett.} 148B, 280.

\bibitem{Vil2} Vilenkin, A. (1984). Quantum creation of universes,\textit{\
Phys. Rev}. D30, 509.

\bibitem{ZelSta} Zeldovich, Ya.B. and Starobinsky, A.A. (1984). Quantum
creation of a universe in an nontrivial topology, \textit{Sov. Astron. Lett.}
10, 135.

\bibitem{HHH1} Hartle, J. B., Hawking S.W. and Hertog T., \textit{Phys. Rev}%
. D 77, 123537 (2008).

\bibitem{HHH2} Halliwell, J. J., Hartle, J. B. and Hertog, T., \textit{Phys.
Rev}. \textbf{D 99}, 043526 (2019).

\bibitem{HH2} Hartle, J. B., Hawking, S. W. and Hertog, T., Quantum
Probabilities for Inflation from Holography, \textit{Journal of Cosmology
and Astroparticle Physics,} 2014(01), p.015 (2014).

\bibitem{H} Hawking, S.W. In: \textit{General Relativity; An Einstein
centenary survey}, ed. By S.W. Hawking, and W. Israel, Cambridge University
press (1980).

\bibitem{Witt} Witten, E., Commun. \textit{Math. Phys}., 80, 381 (1981).

\bibitem{Fadd} Faddeev, L.D., \textit{UFN}, V.136, No, pp.435-457 (1982).

\bibitem{GL} Gorobey, N. N. and Lukyanenko, A.S., \textit{TMF}, V.95, No 3,
pp.541-548 (1992).

\bibitem{MTW} Misner, C. W., Thorne, K. S. and Wheeler, J. A., \textit{%
Gravitation}, W. H. Freeman and Company, San Francisco (1973).

\bibitem{GLG} Gorobey, N., Lukyanenko, A., Goltsev, A. V. Wave Functional of
the Universe and Time. \textit{Universe} 2021, 7, 452-461.

\bibitem{FradVil} Fradkin, E. S.; Vilkovisky, G. A. \textit{Phys. Lett.}
1975, 55B, 224.

\bibitem{BatVil} Batalin, I. A.; Vilkovisky, G. A. \textit{Phys. Lett}. 69B,
309. (1977).

\bibitem{Gov} Govaerts, J. \textit{Int. J. of Modern Physics A} 1989, 4(17),
4487.

\bibitem{ADM} Arnowitt, R.; Deser, S.; Misner, C.W. The dynamics of general
relativity, \textit{In: Gravitation: An Introduction to Current Research,}
ed. Witten L.,Wiley, New-York, p. 227 (1962).

\bibitem{Ash1} Ashtekar, A., \textit{Phys. Rev.} D36, 1587 (1987).

\bibitem{Ash2} Ashtekar, A., \textit{Physica (Utrecgt) }A, 124, 51 (1984).
124, 51 (1984).

\bibitem{GLG1} Gorobey, N. N.; Lukyanenko, A. S.; Goltsev, A. V. \textit{%
Universe }2022, 8, 568.

\bibitem{Ash3} Ashtekar, A., Pawlowski, T. and Singh, P., \textit{Phys. Rev.
Lett.} 96, 141301 (2006).
\end{thebibliography}
\end{document}